\documentclass[conference]{IEEEtran}

\usepackage{array} 
\usepackage{makecell} 
\usepackage{csquotes}

\IEEEoverridecommandlockouts

\usepackage{cite} 
\usepackage{amsmath,amssymb,amsfonts}
\usepackage{algorithmic}
\usepackage{graphicx}
\usepackage{textcomp}
\usepackage{xcolor} 
\usepackage[left=1in,right=1in,top=1in,bottom=1in]{geometry}
\usepackage{mdframed}
\usepackage{hyperref}
\usepackage{subcaption}

\usepackage[colorinlistoftodos]{todonotes}

\def\BibTeX{{\rm B\kern-.05em{\sc i\kern-.025em b}\kern-.08em
    T\kern-.1667em\lower.7ex\hbox{E}\kern-.125emX}}

\begin{document}

\title{\textbf{\huge{Design and Feasibility of a Community Motorcycle Ambulance System in the Philippines}}\\
}

\author{ 

\textbf{Aaron Rodriguez}$^a$,
\textbf{Aidan Chen}$^b$,
\textbf{Ryan Rodriguez}$^c$\\ \\

$^a$VSee, Washington D.C., $^b$VSee, San Jose\\$^c$Department of Industrial and Systems Engineering, Georgia Institute of Technology
}

\maketitle

\begin{abstract}
This study investigates the potential for motorcycle ambulance (motorlance) deployment in Metro Manila and Iloilo City to improve emergency medical care in high-traffic, underserved regions of the Philippines. VSee, a humanitarian technology company, has organized numerous free clinics in the Philippines and identified a critical need for improved emergency services. Motorlances offer a fast, affordable alternative to traditional ambulances, particularly in congested urban settings and remote rural locations. Pilot programs in Malawi, Thailand, and Iran have demonstrated significant improvements in response times and cost-efficiency with motorlance systems. This study presents a framework for motorlance operation and identifies three potential pilot locations: Mandaluyong, Smokey Mountain, and Iloilo City. Site visits, driver interviews, and user surveys indicate public trust in the motorlance concept and positive reception to potential motorlance deployment. Cost analysis verifies the financial feasibility of motorlance systems. Future work will focus on implementing a physical pilot in Mandaluyong, with the aim of expanding service to similar regions contingent on the Mandaluyong pilot's success.
\end{abstract}

\section{Introduction and Background} \label{introduction}

VSee is a medical technology company that develops telemedicine software and organizes humanitarian missions to deliver medical care in low-resource regions.  The VSee humanitarian team has organized hundreds of free clinics in dozens of countries, including many for underserved communities in the Philippines.  A prior VSee air quality study in Metro Manila uncovered the need for improved emergency medical care in high-traffic, low-resource areas in the Philippines \cite{vsee2024sensors}. This study explores the potential for motorcycle ambulance deployment in Metro Manila and Iloilo City to provide emergency care improvements over traditional ambulance services.

Motorcycle ambulances, or motorlances, are an emerging alternative to traditional ambulances.  In regions where operating traditional ambulances is prohibitively expensive and burdened by heavy traffic, motorlances provide a fast, affordable alternative to deliver emergency care. Motorlance pilot programs in Mangochi, Malawi; Khon Kaen, Thailand, and Shiraz, Iran have demonstrated motorlance systems can offer significant cost and time improvements over traditional ambulances while still providing a comparable standard of care \cite{malawimotorlance} \cite{thailandmotorlance} \cite{iranmotorlance}.

In the Malawi case study, motorlances reduced travel time to and from medical facilities, with the average motorlance trip time 35\% to 76\% shorter than the existing ambulance system \cite{malawimotorlance}.  Operating costs were 74\% cheaper than the current system, including the purchase and shipping costs for motorlance vehicles, providing resistance to surges in demand for emergency care \cite{malawimotorlance}. These improvements led to better maternal and neonatal outcomes for patients in the Mangochi region, demonstrating the potential of motorlances to save lives in resource-limited settings.

In the fourth largest city in Thailand, Khon Kaen, a motorlance pilot displayed similar results. Motorlances improved response times by 33\%, particularly during peak traffic hours, and reduced environmental impact by lowering fuel consumption and emissions \cite{thailandmotorlance}. Thailand's strict regulations on required equipment carried by emergency response vehicles may have contributed to slower response times, but nevertheless demonstrated the viability of motorlances in dense urban settings and their ability to deliver emergency medical equipment.

The efficiency of motorlances compared to traditional ambulances was also evaluated in Shiraz, Iran. The study revealed motorlances were able to navigate through congested urban areas more effectively, leading to 50\% faster response times and similar patient outcomes \cite{iranmotorlance}. Consistent with the Malawi study, the Shiraz case study's cost analysis indicated motorlances were far more economical to operate and maintain, making motorlances a viable option for low-income areas with significant budget constraints. The success of the motorlance program in Shiraz has prompted further exploration of its potential applications in other regions of Iran, highlighting the maintainability and scalability of motorlance programs \cite{iranmotorlance}.

The population of the Philippines is rapidly increasing, and city centers are experiencing worsening congestion. Extreme population density, narrow city streets, and high personal vehicle ownership make car navigation difficult, hindering ambulance services. Metro Manila in particular has experienced many of these adverse effects, and the success of motorlance programs in similar cities merits motorlance feasibility assessments for use in the Philippines.

This paper proposes a framework for motorlance operation with continuous rider-dispatcher communication (Section \ref{systemoverview}), identifies locations for potential pilot programs (Section \ref{pilotlocations}), details the results of local driver interviews (Section \ref{driverinterviews}), and analyzes responses from a 100 respondent survey carried out in the proposed pilot region (Section \ref{usersurvey}). A joint VSee and Georgia Institute of Technology team conducted site visits to Mandaluyong, Smokey Mountain, and Iloilo City, Philippines to interview drivers, distribute user surveys, and finalize the location for a future motorlance pilot. Results show promise for motorlance deployment, with strong public approval and positive feasibility assessments from driver interviews.

\section{Overall System} \label{systemoverview}

This study proposes an on-demand transportation dispatch system similar to the design proposed by Van Hentenryck et al. \cite{vanhentenryck2023martareachpilotingondemand} with the addition of rider-dispatcher communication.

From a rider perspective, users interface with the system through a mobile phone app.  In an emergency, users can press a one-tap button to call a motorlance by initiating a call with the dispatcher.  Users can specify the nature of the emergency and provide necessary details either through the app or during the call.  Users track the motorlance in real time and receive updates on its estimated time of arrival.

When users first download the app, they will have the option to register with their personal details, including brief medical history (e.g. asthma, heart problems) and emergency contacts.  Upon calling a motorlance, the backend will automatically notify emergency contacts via message.  The one-tap call button remains available throughout the registration process, ensuring users are not required to register during an emergency.

\begin{figure}[!ht]
     \centering
     \includegraphics[width=1\linewidth]{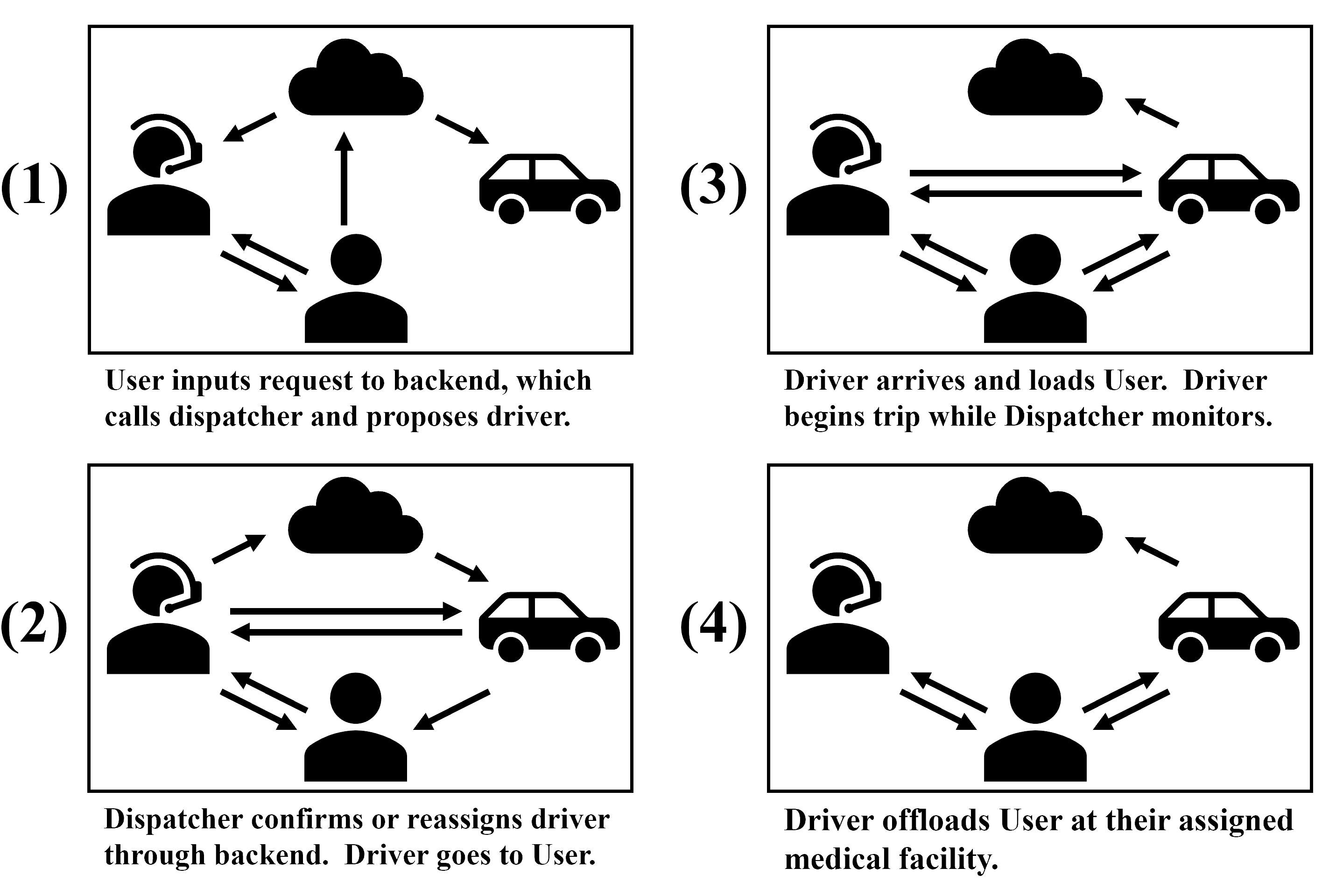}
     \caption{Proposed motorlance system framework.
    }
     \label{motorlanceframework}
 \end{figure}

From a driver perspective, drivers will be assigned a user location and dispatcher.  After receiving their assignment, drivers will drive towards the user location, receiving guidance from their assigned dispatcher along the way.  Drivers undergo rigorous training in initial medical assistance to allow for immediate emergency response upon arrival.  Users will be safely loaded into the motorlance and driven to their assigned medical facility.

From a dispatcher perspective, dispatchers receive initial alerts when a user opens the app.  When an emergency request is placed, the backend server selects the closest available driver to serve the request.  This choice is displayed to the dispatcher, who has a short period to respond by confirming the current driver or reassigning the request to another driver. Since the dispatcher immediately receives a call from the user upon receiving the request, the delay period allows the dispatcher to assess the situation and coordinate with the driver as needed.  Dispatchers monitor users and drivers until service is completed, coordinating with the medical facility as needed.

Drivers will not be notified of their assignment until assigned by a dispatcher or the delay period expires.  Users will be automatically assigned a medical facility that can be easily changed by the dispatcher.

While drivers can be automatically dispatched if a dispatcher is not present, user surveys (see Section \ref{usersurvey}) and prior literature indicate human dispatchers increase trust and usage of transportation systems \cite{vanhentenryck2023martareachpilotingondemand}.  The use of human dispatchers is also advantageous for liability purposes, as requests that cannot be serviced by motorlance can be easily reassigned to existing emergency services.

Drivers and dispatchers must pass a screening process before serving users.  Unlike traditional ridehailing apps, both will undergo a background check and be directly hired for stable employment in motorlance operation, as opposed to the tenuous "at-will" employment of transportation network companies.

\section{Methodology}

\subsection{Potential Pilot Locations}  \label{pilotlocations}

As part of its humanitarian initiatives, VSee has operated free health clinics in Metro Manila for nearly a decade \cite{pickering2024building}.  Potential motorlance pilot locations were identified by VSee contacts in the Philippines and narrowed down to three finalists: Smokey Mountain, Mandaluyong, and Iloilo City.

\begin{figure}[!ht]
     \centering
     \includegraphics[width=1\linewidth]{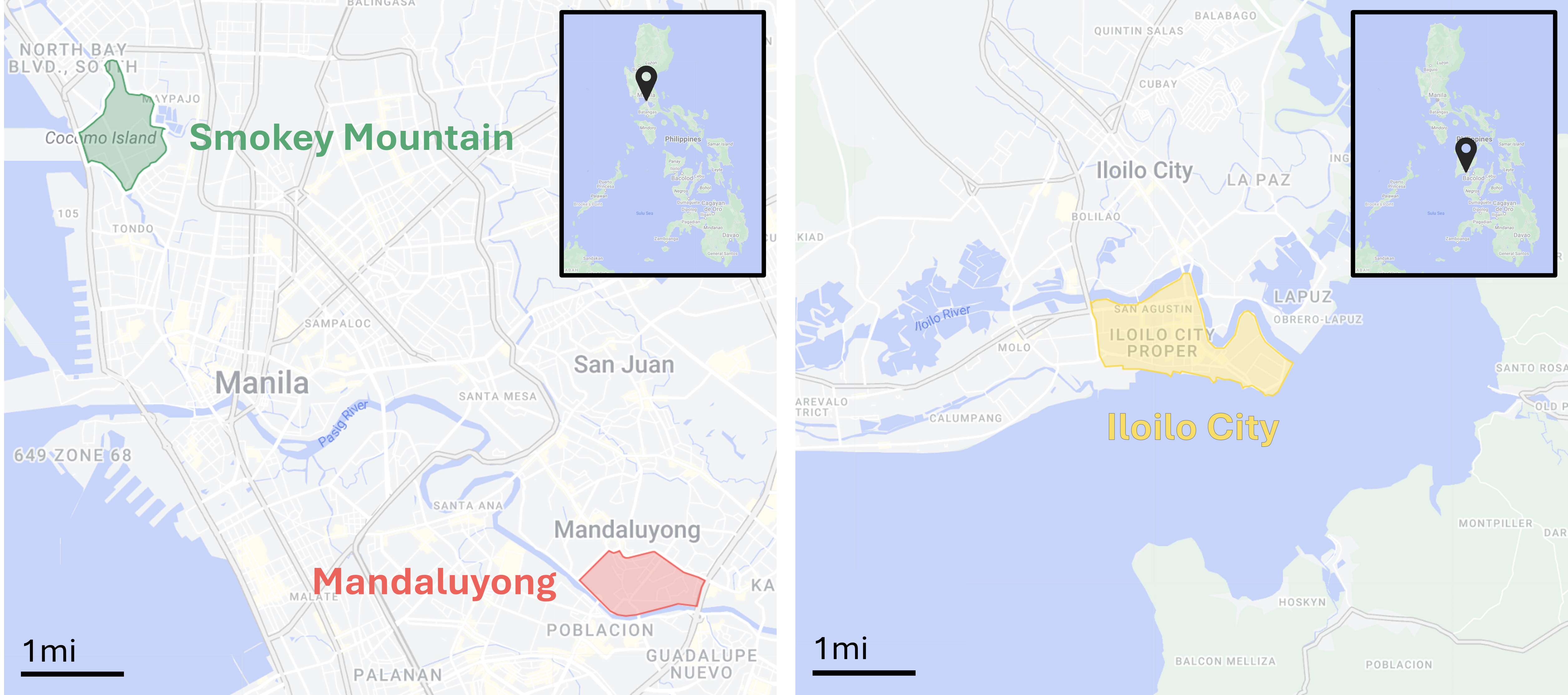}
     \caption{Potential pilot regions: Smokey Mountain (green), Mandaluyong (red), and Iloilo City (yellow).  Minimap in top right displays location relative to rest of Philippines.
    }
     \label{pilotregions}
 \end{figure}

\begin{figure*}[!ht]
\centering
\includegraphics[height=11.5em]{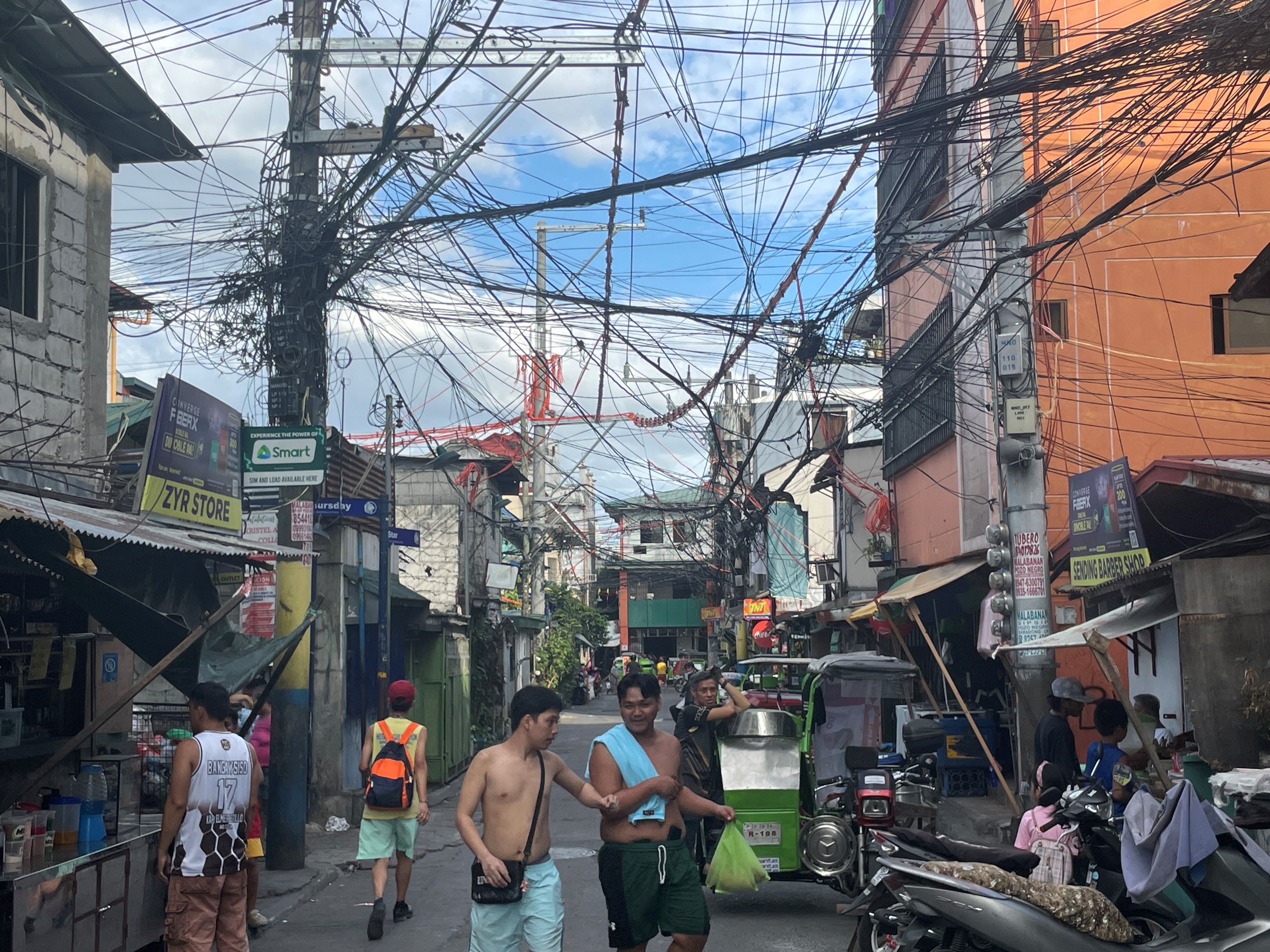}
\includegraphics[height=11.5em]{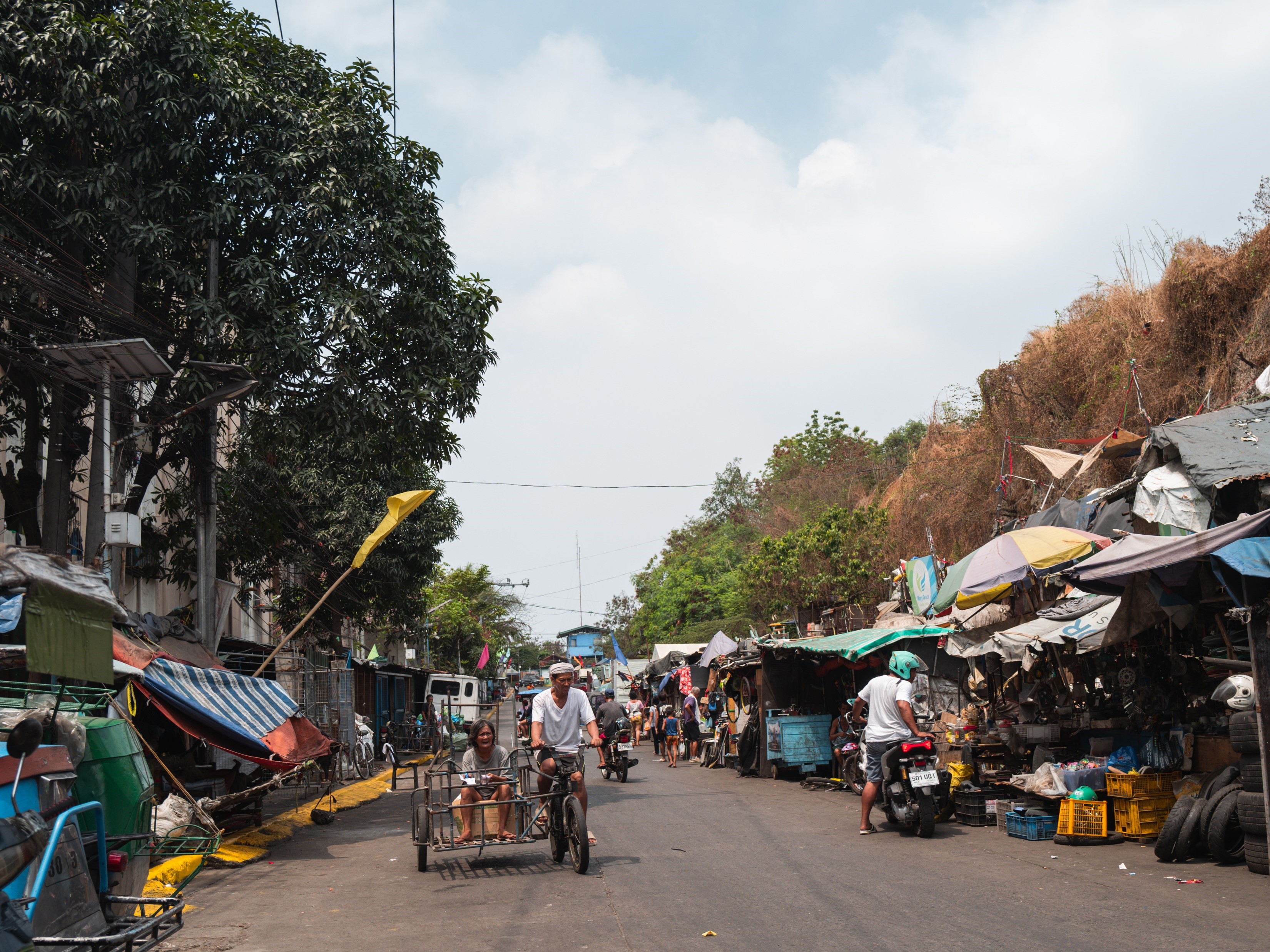}
\includegraphics[height=11.5em]{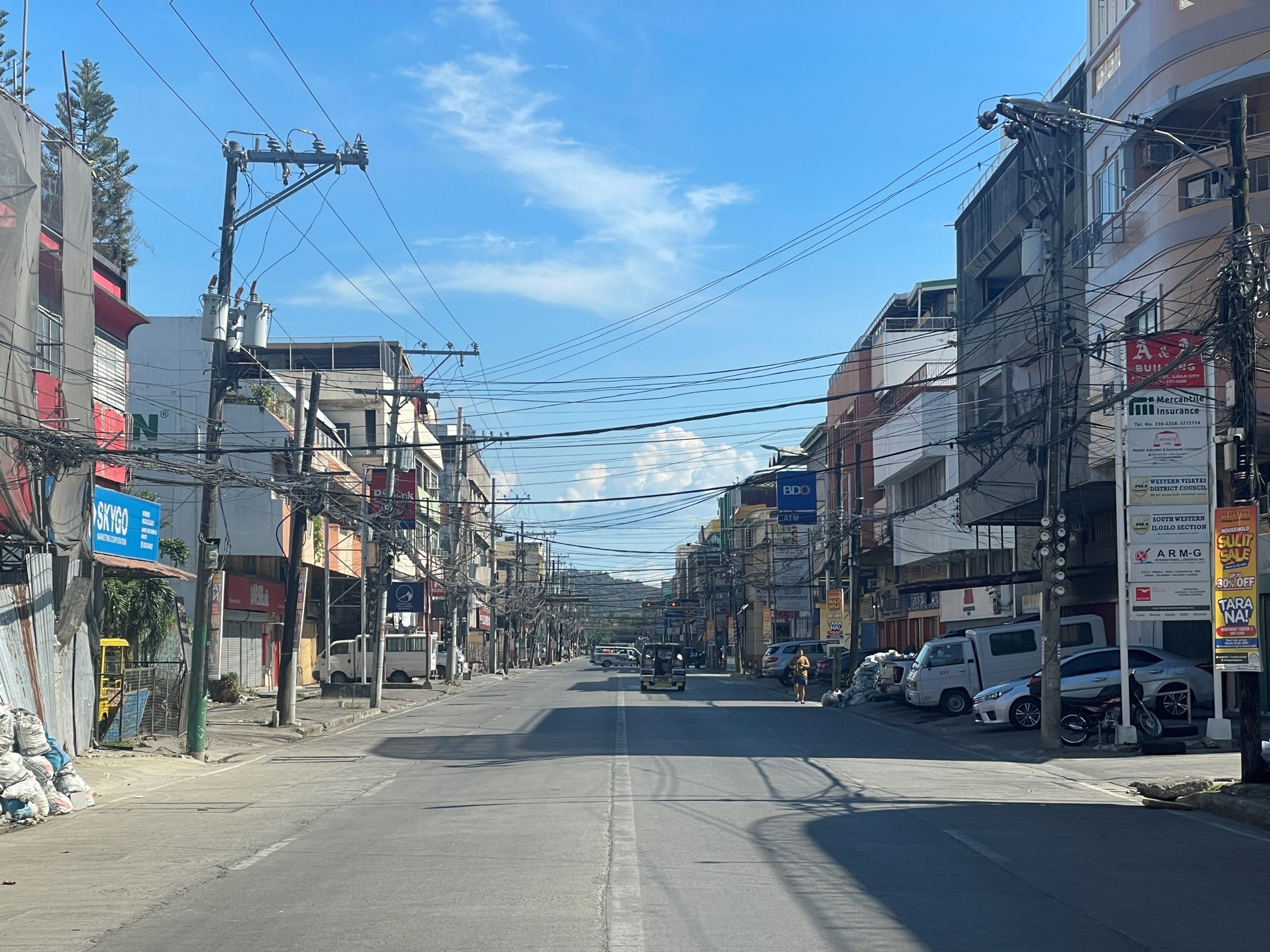}
\captionsetup{justification=centering}
\caption{On-site street conditions for Mandaluyong (left), Smokey Mountain (center), and Iloilo City (right).}
\label{streetviews}
\end{figure*}

Smokey Mountain was selected due to the proximity of a large landfill that gives the region its name.  Smokey Mountain residents experience poverty and pollution far above the national average, with the main road running along the side of the landfill \cite{abad1991squatting}.  Poor streetside conditions (see Figure \ref{streetviews}), low-income population, and pollutant hazards indicate residents may experience elevated health risks, making Smokey Mountain a viable pilot location \cite{vsee2024sensors}.

Mandaluyong was selected because of its widespread internet access, generally favorable outlook on technology, and unusually high congestion due to narrow street design.  Technology-focused educational institutions like Rizal Technical University and Polytechnic University of the Philippines are located in Mandaluyong, and VSee regularly operates free clinics at the City of Mandaluyong Science High School \cite{vsee2024sensors}.

With the other two locations in Metro Manila, Iloilo City was selected to assess conditions outside the Philippines' highly congested capital.  Traffic flow studies indicated both that Iloilo City would experience increased congestion in the near future and that Iloilo residents were likely to consider mode switching if offered improvements in travel time, ideal conditions for motorlance operation \cite{sosuan2014mode}.

\subsection{Driver Interviews} \label{driverinterviews}

To obtain local, professional perspectives on motorlance feasibility, three transport network company drivers were interviewed on traffic conditions and motorlance viability.  Drivers were selected from Mandaluyong (\textbf{Driver 1}), Iloilo City (\textbf{Driver 2}), and Manila proper (\textbf{Driver 3}), corresponding to the three potential pilot locations discussed previously.

Short interviews with each driver were recorded and transcribed verbatim, including all speech irregularities.  Excerpts from these interviews are displayed below, with notes to define terms or provide context unitalicized and in parenthesis.  Full transcripts can be provided upon request.

Overall, drivers corroborated many of the study's initial assumptions (see Section \ref{pilotlocations}).  All three drivers indicated local government was responsible for financing and operating traditional ambulance services, and smaller vehicles like motorcycles or tricycles (motorbikes with sidecars) can get around faster than larger vehicles like jeepneys (elongated jeeps used as buses):

\begin{displayquote}
\textit{\\
\textbf{Interviewer:} Let’s say someone is here in this building and they hurt their arm. How do they get a hold of the… how do they call the ambulance?\\
\textbf{Driver 1:} I think that uh, that they call the barangay} (small administrative district, similar to a U.S. census tract) \textit{and they call the 911.\\
\textbf{Interviewer:} So 911?\\
\textbf{Driver 1:} 911 and the barangay is for the ambulance.\\}
\end{displayquote}

\begin{displayquote}
\textit{\textbf{Interviewer:} So if you have somebody in your in your home who has a heart problem, how do how do you call the ambulance?\\
\textbf{Driver 3:} Uh… we can call uh, first you can call the barangay and then the barangay can take- you can call them sir.\\
}
\end{displayquote}

Timeliness of emergency medical services is also a clear challenge, an area motorlances can significantly improve on:

\begin{displayquote}
\textit{\\
\textbf{Interviewer:} How long does it take for them} (ambulances) \textit{to come here?\\
\textbf{Driver 1:} Uh, almost uh, 15 minutes. How is very long. Usually, in case of fire, uh, very long. The fireman, the firetruck will arrive} (unclear) \textit{in 30 minutes. The house or the building was burnt.}
\end{displayquote}

\begin{displayquote}
\textit{\\\textbf{Interviewer:} So you call the barangay and they will call somebody, the ambulance to come out to get you?\\
\textbf{Driver 3:} Yeah…\\
\textbf{Interviewer:} How long does that take?\\
\textbf{Driver 3:} It’s either… it depends sir, uh, to how long for example it have been sir.  It’s very, very long ride sir.\\
}
\end{displayquote}

Driver 1 reported motorcycles could reach locations five times faster than traditional ambulances, and three times faster with a motorlance-like attachment (e.g. a sidecar):

\begin{displayquote}
\textit{\\
\textbf{Interviewer:} Do ambulances come reliably or no?\\
\textbf{Driver 1:} Yeah it’s uh, sometimes we feel sad because uh, very slow action, especially the government.\\
\textbf{Interviewer:} Gotcha. Do you think motorcycles would be able to… like a motorcycle or moped would be able to be used like as an ambulance faster?\\
\textbf{Driver 1:} Yeah, the motorcycle is moving faster here.\\
\textbf{Interviewer:} So if the ambulance takes 15 minutes how fast could the motorcycle get there?\\
\textbf{Driver 1:} Uh, the motorcycle get there is almost a total 3 minutes. The motorcycle is uh…} [trails off]\\
\textit{\textbf{Interviewer:} How about a sidecar?\\
\textbf{Driver 1:} If you have a sidecar almost about five minutes.\\
}
\end{displayquote}

Surprisingly, while both Manila-based drivers (Drivers 1 and 3) validated initial research reporting heavy traffic and non-ideal road conditions, Iloilo-based Driver 2 described few issues:

\begin{displayquote}
\textit{\\
\textbf{Interviewer:} Does it take a long time to get the ambulance or no?\\
\textbf{Driver 2:} It depends on the area sir.  It depends on how far of your house in the city.\\
\textbf{Interviewer:} What happens if you live far away from the city, what happens if you live north in the island, is it harder to get the ambulance or no?\\
\textbf{Driver 2:} Uh… no sir, because every barangay they have one ambulance.\\
\textbf{Interviewer:} And so… is traffic so bad that it’s hard to get in with the ambulance, or no?\\
\textbf{Driver 2:} Uh, no, sir. Because uh, here in Iloilo sir, they uh, they have no traffic.\\
\textbf{Interviewer:} I see.\\
\textbf{Driver 2:} Yes. Just like- unlike Manila.\\}
\end{displayquote}

Due to Iloilo's developed road infrastructure and limited traffic, Driver 2 suggested motorlances would be less feasible for its developed, non-congested setting:

\begin{displayquote}
\textit{
\textbf{Interviewer:} And the roads are, the roads are pretty wide I see, so-\\
\textbf{Driver 2:} Yes.\\
\textbf{Interviewer:} -trucks and cars can go pretty much anywhere? Or cars can go pretty much anywhere instead of motor bikes?\\
\textbf{Driver 2:} Yes, because here in Iloilo they have lots of uh, special road and the bypass road.\\
}
\end{displayquote}

Indeed, the Iloilo City site visit supported this view, with wide roads and limited congestion on most city streets (see Figure \ref{streetviews}). However, despite Iloilo City's non-suitability for motorlance deployment, it is important to continue to consider motorlance viability in more removed rural settings.  As seen in the Malawi case study, motorlances can expand service to areas inaccessible by traditional ambulance systems, such as remote unpaved roads \cite{malawimotorlance}.  Future deployments should consider this additional use case when making deployment location decisions.

\subsection{Survey Distribution} \label{usersurvey}

Community perception and feedback on motorlance deployment is crucial to the successful implementation of these systems.  An 11 question user survey in English (with translations in Tagalog) was developed to assess accessibility to and public perception of a potential motorlance pilot program:

\begin{enumerate}
\item[\textbf{Q1}] Age \textit{(text entry)}
\item[\textbf{Q2}] Sex \textit{(male, female)}
\item[\textbf{Q3}] Do you have a university or college degree? \\ \textit{(yes, no)}
\item[\textbf{Q4}] Do you have reliable internet or cell service? \\ \textit{(yes, no)}
\item[\textbf{Q5}] Do you have a cell phone? \textit{(yes, no)}
\item[\textbf{Q6}] If so, what brand (e.g. Samsung, Oppo)? \\ \textit{(text entry)}
\item[\textbf{Q7}] Circle how often do you use transportation apps like Grab, where 1 is "never" and 5 is "very often". \textit{(five point Likert scale)}
\item[\textbf{Q8}] "Imagine a free motorcycle or Jeepney ambulance service where you can call the ambulance to get to the hospital using an app." How likely would you be to use the described system, where 1 is "never" and 5 is "very likely"? \textit{(five point Likert scale)}
\item[\textbf{Q9}] Would you trust a system like this? \\ \textit{(five point Likert scale)}
\item[\textbf{Q10}] What would you currently use if an emergency occurred? \textit{(text entry)}
\item[\textbf{Q11}] \{Optional\} What features would you like to see for the system described? \textit{(text entry)}
\end{enumerate}

Due to limited resources, surveying all three potential pilot locations was unfeasible.  Mandaluyong was selected for surveying due to driver interviews and congestion conditions suggesting an ideal environment for a pilot.

100 copies of the survey were distributed via ground around southern Mandaluyong, near the City of Mandaluyong Science High School.  This portion of Mandaluyong was selected due to existing VSee contacts and the prevalence of narrow street design mentioned previously.  Tagalog responses to the two questions requiring text input were directly translated to English for data processing, and responses missing 2 or more fields were excluded from analysis.  $n = 96$ responses were used in survey analysis.

Collected responses displayed slight sampling bias towards young adults and women, with 57 (60\%) female respondents (\textbf{Q2}) and almost all (98.9\%) respondents younger than 55 (\textbf{Q1}). The slight response bias displayed is due in part to the region of Mandaluyong where the survey was conducted, and results are expected to be representative of the area's population. Full survey data is available upon request from the corresponding author.

\begin{figure}[!ht]
     \centering
     \includegraphics[width=1\linewidth]{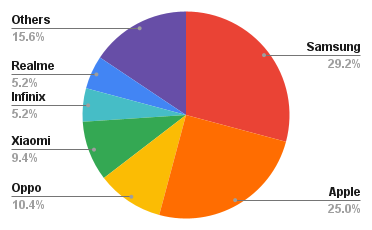}
     \caption{Primary phone brands (\textbf{Q6}) from the survey.
    }
     \label{phone_pie}
 \end{figure}

User accessibility is a primary consideration for app development. Almost all respondents indicated the ability to use a motorlance app, with 100\% (96) of respondents owning a cell phone (\textbf{Q5}) and 98.9\% (95) reporting reliable internet or cell service (\textbf{Q4}).  The vast majority of users reported owning a phone (see Figure \ref{phone_pie}) with an iOS or Android based operating system, with the notable exception of 3 Huawei users, who may be using HarmonyOS.  While the majority of users surveyed use Android based operating systems, a significant percentage of users (25\%) use Apple's iOS.  This suggests a motorlance app should be developed for both the Google Play Store and Apple App Store to maximize user base.

\begin{figure}[!ht]
     \centering
     \includegraphics[width=1\linewidth]{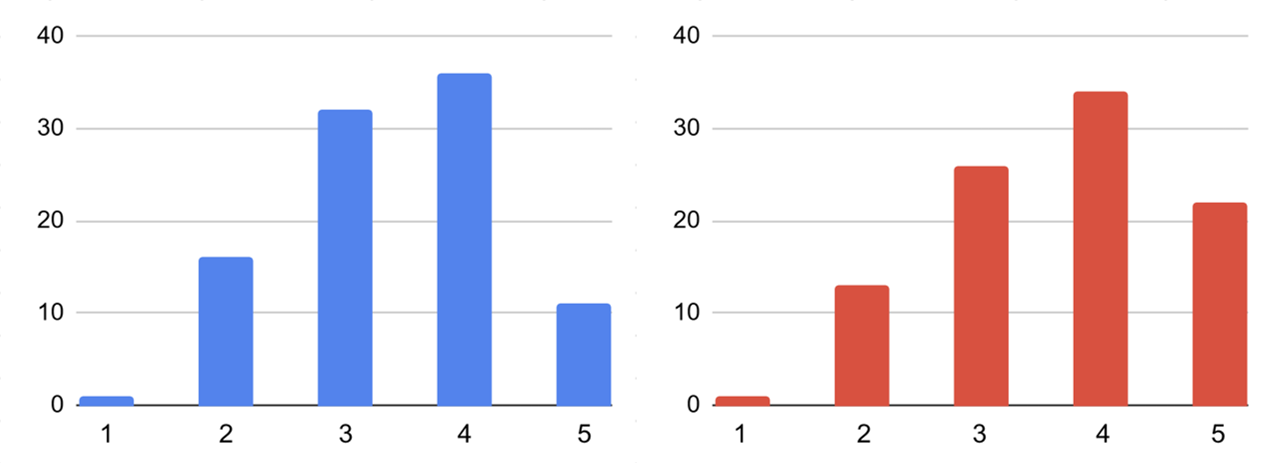}
     \caption{Likert scale responses for system trust (\textbf{Q9}, left) and likeliness to use (\textbf{Q8}, right).
    }
     \label{trust_likely}
 \end{figure}

\begin{figure}[!ht]
     \centering
     \includegraphics[width=1\linewidth]{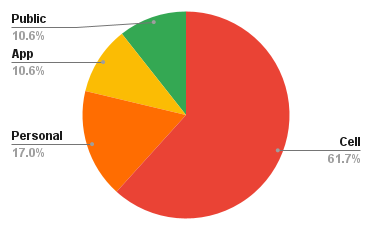}
     \caption{Preferred \underline{existing} system to arrange emergency care (\textbf{Q10}).
    }
     \label{emergency}
 \end{figure}

Questions about the motorlance concept enjoyed relatively high levels of trust (\textbf{Q9}) and potential usage (\textbf{Q8}), as shown in Figure \ref{trust_likely}. Many users selected similar scores for likeliness to use and system trust, implying trust and familiarity with a motorlance system can be a barrier to adoption.  Fortunately, roughly half (46 respondents, or 47.9\%) of those surveyed report regularly using transportation apps (\textbf{Q7}), suggesting widespread familiarity with app-based transportation systems.  Furthermore, the surveyed population appears receptive to an app-based system for emergency care, with 10.6\% of respondents already using a transportation network company app in the event of medical emergency (see Figure \ref{emergency}). 

Surprisingly, younger respondents tended to favor phone based systems. When asked what existing system they would use to arrange emergency care (\textbf{Q10}), 4 of the 10 oldest respondents indicated they would primarily use a non-call based system (e.g. personal vehicles or transportation network companies), while 7 of the 10 youngest indicated they would use a call-based system.  These results reinforce the need for built-in cell communication between users and dispatchers.  Indeed, this trend was observed in responses to \textbf{Q11} (desired features of a motorlance system):

\begin{displayquote}
\textit{\\
\textbf{Response \#33:} "Healthcare numbers."\\
\textbf{Response \#70:} "Safe call number."\\
\textbf{Response \#70:} "Easy way to call and one tap button to call the emergency ambulance."\\
}
\end{displayquote}

Respondents also indicated interest in real-time tracking, ease of use, and security:

\begin{displayquote}
\textit{\\
\textbf{Response \#1:} "Actual / Live location of the ambulance."\\
\textbf{Response \#43:} "One tap emergency response."\\
\textbf{Response \#76:} "To be officially checked by officials."\\
}
\end{displayquote}

Users that displayed low levels of trust and likeliness to use a motorlance system in \textbf{Q8} and \textbf{Q9} also tended to express concerns in \textbf{Q11} about government approval, suggesting rider adoption will increase if the system's collaboration with government officials is advertised.

\subsection{Cost Analysis}

The financial feasibility of replacing traditional ambulances with motorlances in Mandaluyong is crucial to the pilot program's success. From the Philippines Department of Health, the procurement cost of ambulances in Metro Manila fully outfitted with basic life support equipment can range from PHP 2,500,000 to PHP 1,500,000 per unit \cite{pna_1152569}. The high cost of traditional ambulance systems has received significant media attention in the early 2020s due to the COVID-19 pandemic \cite{pna_1152569}.

\begin{table}[!ht]
\begin{tabular}{lllll}
           & \multicolumn{2}{l}{\begin{tabular}[c]{@{}l@{}}Expected Procurement and\\ Outfitting Cost (PHP)\end{tabular}} & \multicolumn{2}{l}{\begin{tabular}[c]{@{}l@{}}Cost Compared to \\ Ambulance Cost (\%)\end{tabular}} \\ \cline{2-5} 
           & Max. & Min. & Max. & Min. \\ \hline
Ambulance  & 2,500,000 & 1,500,000 & 100\% & 100\% \\
Motorlance & 262,500 & 112,500 & 17.5\% & 4.5\% \\ \hline       
\end{tabular}
\caption{Cost analysis of ambulance and motorlance vehicles.}
\label{tab:costanalysis}
\end{table}

The procurement cost of motorlances are significantly cheaper, ranging from PHP 150,000 to PHP 75,000 per unit \cite{hondamotors}. Due to the novelty of motorlance systems, no estimates could be obtained for motorlances outfitted with basic life support equipment, but a 50-75\% cost increase aligns with the Malawi case study \cite{malawimotorlance}, bringing the total procurement and outfitting cost range from PHP 262,500 to PHP 112,500.

The stark difference in initial investment cost displayed in Table \ref{tab:costanalysis} highlights the affordability of motorlance systems.  With the price for motorlance procurement and outfitting ranging from 4.5-17.5\% of the cost for a traditional ambulance, multiple motorlances can be acquired for the cost of a single ambulance. Similar cost reductions were observed for the Malawi \cite{malawimotorlance} and Shiraz \cite{iranmotorlance} case studies mentioned in Section \ref{introduction}.

In addition to lower acquisition costs, motorlances offer  operational savings due to their lower maintenance and fuel expenses, making them a cost-effective solution for providing timely medical response in urban areas like Mandaluyong. By reallocating funds from expensive traditional ambulances to a lightweight motorlance fleet, cities can enhance their emergency response capabilities while staying within budget constraints.

\subsection{Government Approval}

Local government approval is necessary for future motorlance pilot programs.  VSee is discussing the possibility of pilot deployment in Mandaluyong with relevant government officials, but cannot provide details at the time of publication.

\section{Conclusions and Next Steps}

This study presents a design framework for a community motorlance system and assesses deployment feasibility for multiple regions in the Philippines. An on-demand transportation dispatch system was outlined for use in a future motorlance pilot program, incorporating user feedback from surveys to improve ease of use. Due to road conditions, driver testimony, survey results, and existing VSee contacts, Mandaluyong was identified as the ideal location for a future pilot. Cost analysis supports motorlance viability in Metro Manila.

Community support is crucial to the successful deployment of future motorcycle ambulance systems. Preliminary survey findings suggest high levels of accessibility and trust among respondents.  Familiarity with existing app-based transportation and high levels of willingness to use further support the feasibility of an app-based motorlance system. The inclusion of features such as real-time tracking, a one button call system, and advertised partnership with local government will enhance user trust and satisfaction.

Future studies aim to further assess motorlance viability through deployment of a physical pilot program in Mandaluyong, with potential expansion to Smokey Mountain and other locations, especially remote rural regions, to evaluate motorlance performance in various contexts. VSee will continue to explore these options alongside existing air quality sensor deployment projects in the Philippines.


\begin{thebibliography}{10}

\bibitem{abad1991squatting}
Ricardo~G Abad.
\newblock Squatting and scavenging in smokey mountain.
\newblock {\em Philippine Studies}, 39(3):263--286, 1991.

\bibitem{thailandmotorlance}
Korakot Apiratwarakul, Takaaki Suzuki, Ismet Celebi, Somsak Tiamkao, Vajarabhongsa Bhudhisawasdi, Chatkhane Pearkao, and Kamonwon Ienghong.
\newblock “motorcycle ambulance” policy to promote health and sustainable development in large cities.
\newblock {\em Prehospital and Disaster Medicine}, 37(1):78--83, 2022.

\bibitem{vsee2024sensors}
Aidan Chen, Jonathan Du, Aaron Rodriguez, Ryan Rodriguez, Jack Higgins, Robin Podmore, Ryan Liu, Emin Ilao, Sam Degilla, James Bibiano, Candice Chang, Annalicia Pickering, Mary Showstark, Jarone Lee, and Milton Chen.
\newblock Viability of applying large language models to indoor climate sensor and health data for scientific discovery.
\newblock {\em IEEE Global Humanitarian Technology Conference}, 14(1), 2024.

\bibitem{vanhentenryck2023martareachpilotingondemand}
Pascal~Van Hentenryck, Connor Riley, Anthony Trasatti, Hongzhao Guan, Tejas Santanam, Jorge~A. Huertas, Kevin Dalmeijer, Kari Watkins, Juwon Drake, and Samson Baskin.
\newblock Marta reach: Piloting an on-demand multimodal transit system in atlanta, 2023.

\bibitem{malawimotorlance}
Jan~J Hofman, Chris Dzimadzi, Kingsley Lungu, Esther~Y Ratsma, and Julia Hussein.
\newblock Motorcycle ambulances for referral of obstetric emergencies in rural malawi: do they reduce delay and what do they cost?
\newblock {\em International Journal of Gynecology \& Obstetrics}, 102(2):191--197, 2008.

\bibitem{iranmotorlance}
Mahmoud~Reza Peyravi, F~Toubaei, and K~Pourmohammadi.
\newblock The efficiency of motorlance in comparison with ambulance in shiraz, southern iran.
\newblock 2009.

\bibitem{pna_1152569}
Philippine News Agency~(GOV PH).
\newblock Department of health defends purchase of medical-grade ambulances, 2023.

\bibitem{hondamotors}
Honda~Motors Philippines.
\newblock Motorcycles for business, 2024.

\bibitem{pickering2024building}
Annalicia Pickering, Wardah Rafaqat, Adi Balk, Milton Chen, Limuel Abrogena, Shuhan He, Mary Showstark, Alexander Davis, Aidan Chen, and Jarone Lee.
\newblock A building blocks approach to implementing a telehealth clinic model to improve primary care access in the philippines: A large-scale pilot project.
\newblock {\em Telehealth and Medicine Today}, 9(1), 2024.

\bibitem{sosuan2014mode}
Frederick Lloyd~A Sosuan and Alexis~M Fillone.
\newblock Mode choice analysis of urban trips in iloilo city.
\newblock In {\em Proceedings of the 22nd Annual Conference of the Transportation Science Society of the Philippines}, 2014.

\end{thebibliography}
\end{document}